\documentclass[article,12pt]{article}
\usepackage{etex}
\usepackage[top=1in,bottom=1in,left=1in,right=1in]{geometry}
\usepackage{slashed}
\usepackage{epsfig}
\usepackage{comment}
\usepackage{cancel}
\usepackage{bbm}
\usepackage{array}
\usepackage{bigints}
\usepackage{braket}
\usepackage{booktabs}
\usepackage{color}
\usepackage{dsfont}
\usepackage{float}
\usepackage{framed}
\usepackage{graphicx}
\usepackage{indentfirst}
\usepackage{mathrsfs}
\usepackage{multirow} 
\usepackage{subdepth}
\usepackage{subfig}
\usepackage{titlesec}
\usepackage[dotinlabels]{titletoc}
\usepackage{wrapfig}
\usepackage[all]{xy}
\usepackage{young}
\usepackage[vcentermath]{youngtab}
\usepackage{relsize}
\usepackage{hyperref}
\usepackage{feynmp}
\usepackage{appendix}
\numberwithin{equation}{section}

\usepackage[utf8]{inputenc}
\usepackage{geometry}
\usepackage{graphicx}
\usepackage{array}
\usepackage{subfig}
\usepackage{slashed}
\usepackage{amsmath}
\usepackage{amsfonts}
\usepackage{amssymb}
\usepackage{cite}
\usepackage{epsfig}
\usepackage{comment}
\usepackage{tikz}
\usetikzlibrary{patterns}
\usepackage{feynmp}
\usepackage{cancel}
\usepackage{mathrsfs}
\usepackage{mathtools}

\usepackage{cleveref}
\def\be{\begin{equation}}
\def\ee{\end{equation}}
\def\bz{{\bar z}}

\def\ba{\bar \alpha}
\def\p{\partial}

\def\be{\begin{equation}}
\def\ee{\end{equation}}
\def\bea{\begin{eqnarray}}
\def\eea{\end{eqnarray}}
\def\<{\langle }
\def\>{\rangle}

\def\SL{\mathrm{SL}}
\def\CC{\mathbb{C}}
\def\RR{\mathbb{R}}

\def\bh{\bar{h}}
\def\d{\partial}
\def\O{\mathcal{O}}
\def\rn{\mathbb{R}}
\def\ints{\mathbb{Z}}

\newcommand*\pFqskip{8mu}
\catcode`,\active
\newcommand*\pFq{\begingroup
        \catcode`\,\active
        \def ,{\mskip\pFqskip\relax}%
        \dopFq
}
\catcode`\,12
\def\dopFq#1#2#3#4#5{%
        {}_{#1}F_{#2}\biggl[\genfrac..{0pt}{}{#3}{#4};#5\biggr]%
        \endgroup
}
\crefname{section}{§}{§§}
\Crefname{section}{§}{§§}

\begin{document}
\begin{titlepage}
\unitlength = 1mm
\ \\
\vskip 3cm
\begin{center}

{\LARGE{\textsc{Conformal Block Expansion in Celestial CFT}}}

\vspace{0.8cm}
Alexander Atanasov$^*$, Walker Melton$^*$, Ana-Maria Raclariu$^\dagger$ \\
and Andrew Strominger$^*$
\vspace{1cm}

$^*${\it  Center for the Fundamental Laws of Nature, Harvard University,\\
Cambridge, MA 02138, USA}\\
$^\dagger${\it Perimeter Institute for Theoretical Physics, Waterloo, ON N2L 2Y5, Canada}

\vspace{0.8cm}

\begin{abstract}

The 4D 4-point scattering amplitude of  massless scalars  via a massive exchange is expressed in a basis of 
conformal primary particle wavefunctions. This celestial amplitude is expanded in a basis of 2D conformal partial waves on the unitary principal series, and then rewritten as a sum over 2D conformal blocks via contour deformation. The conformal blocks  include
intermediate exchanges of spinning light-ray states, as well as scalar states  with positive integer conformal weights.  
The conformal block  prefactors are found as expected to be  quadratic in the  celestial OPE coefficients. 

 \end{abstract}

\vspace{1.0cm}

\end{center}

\end{titlepage}

\pagestyle{empty}
\pagestyle{plain}
\def\vx{{\vec x}}
\def\p{\partial}
\def\po{$\cal P_O$}
\def\cb{N_{gm}(\beta)}

\pagenumbering{arabic}
 

\tableofcontents
\newpage
\section{Introduction}

Celestial conformal field theories (CCFTs) provide potential holographic duals for 4D quantum theories of  gravity in asymptotically flat spacetimes \cite{Pasterski:2016qvg,Pasterski:2017kqt,Stieberger:2018edy,Pate:2019mfs,Puhm:2019zbl,Pate:2019lpp,Fotopoulos:2019vac,Banerjee:2020kaa,Fotopoulos:2020bqj,Guevara:2021abz}. Correlators on the 2D celestial sphere are defined as spacetime scattering amplitudes of particles with $SO(3,1)$ conformal primary wavefunctions.  These celestial correlators have some, but not all, of the properties of  standard CFT$_2$ correlators.  In order to better understand their properties,  in this paper we derive a conformal block expansion of the massless scalar 4-point correlators arising from massive scalar exchange. Our  formula reveals a number of interesting features and provides insight into the structure of  CCFT. 
The expansion is 
a  new decomposition of   spacetime scattering amplitudes as  sums over $s$, $t$ OR $u$ channel exchanges, unlike the Feynman diagrammatic decomposition involving $s$, $t$ AND $u$ exchanges. 
This is one of many instances in which stringy behavior emerges in celestial holography. 

One motivation for our  analysis is to find a suitable basis of primary operators in CCFT. 
Depending on its presentation, the   problem of finding a complete basis of primaries of any  conformal field theory  can be quite nontrivial. 
For example in Liouville theory, presented as an action for a scalar field, exponentials of the Liouville field formally describe primaries with arbitrary conformal weight $\Delta$. However a complete basis includes only primaries with $\Delta>{c-1 \over 12}$ on the real axis \cite{Dorn:1994xn,Zamolodchikov:1995aa}. This conclusion requires analysis of modular properties, OPE locality and closure  and crossing symmetry of the 4-point function.  In sine-Liouville theory the situation is even more subtle and some discrete 
states are also required \cite{Hanany:2002ev}. 

The presentations of CCFTs given so far\footnote{An intrinsic construction of any CCFT starting from a microscopic theory of gravity such as string theory would be major progress.} are novel:  2D correlators are defined as Mellin transforms\footnote{This paper considers  massless scattering only. In the massive case a hyperbolic transform replaces the Mellin transform.} of 4D
momentum-space scattering amplitudes. Soft theorems \cite{Kapec:2014opa,Kapec:2016jld,Fotopoulos:2019tpe} imply that these transform as conformal primaries under the Virasoro action on the celestial sphere.\footnote{This only holds when gravity is included, but we nevertheless expect the pure field theory correlators discussed here to capture some aspects of conformal symmetry, particularly the global Lorentz subgroup of Virasoro.} The Mellin transform formally produces primaries of any complex conformal weight. Yet further primaries are obtained from these via shadow or light-ray transforms.  All of these together are clearly an overcomplete basis. An important constraint on a complete basis is that it should supply all the intermediate exchange channels represented in the conformal block expansion of the 4-point correlators.  Hence the present work sheds some light on this open problem. 

Pasterski and Shao \cite{Pasterski:2017kqt} showed - up to zero mode subtleties \cite{Donnay:2018neh} - that states on the unitary principal series with $\Delta=1+i\lambda, ~~\lambda \in {\mathbb{R}}$ form  a complete basis for normalizable (massless) particle scattering states. However such states are not in representations of the translation group, which shift $\Delta \to \Delta +1$, and in gauge or gravity theory they do not include the states 
at non-positive integer $\Delta$ which generate the soft symmetries \cite{Fan:2019emx, Pate:2019mfs,Puhm:2019zbl,Guevara:2021abz}  or their Goldstone boson partners at positive integer $\Delta$.
States with real integral $\Delta$ do include these states and provide representations of the symmetry groups. Moreover the poles in their scattering amplitudes encode all information about the Wilsonian effective action \cite{Arkani-Hamed:2020gyp}. However  it is not known if the integer-weight states  are in any sense complete. 

Our analysis is simplified by working in $(2,2)$ signature Klein space which means the CCFT lives in $(1,1)$ signature \cite{Atanasov:2021oyu}, and by defining correlators with general complex conformal weights via  analytic continuation. Our explicit formulae decompose 4-point scattering of massless scalars via a contact interaction or massive t-channel scalar exchange. All of the intermediate  conformal blocks exchanged can be identified as one of two types of conformal primary states of the massive scalar.  The first type are
spin $J=0$ blocks with real, positive, integral conformal weights $\Delta=2,3,\cdots$.\footnote{The full 4-point scattering mediated by massive exchange also includes spinning exchanges of dimensions $\beta/2$ shifted by integers.}  The second are the so called light-ray states 
which have appeared in a number of recent conformal block expansions \cite{Caron-Huot:2017vep,Kravchuk:2018htv,Kologlu:2019mfz}. They have imaginary spin (as allowed in $(1,1)$)
and $\Delta=1$. They are related by the light-ray transform, which is a shadow in one null direction, to ordinary primaries on the unitary principal series. New conformal primary solutions to the bulk scalar wave equation are herein constructed which describe the wave functions of the associated spacetime scalar particles. 

Conformal symmetry predicts that the coefficients of the conformal blocks are quadratic in the OPE coefficients. The OPE coefficients in CCFT can be extracted from the 3-point amplitude and we indeed find this to be the case. For the integer
blocks the relevant OPEs are Euler beta functions with arguments related to the conformal weights, while for the light-ray transforms they are just the scalar coupling  in an appropriate basis. 

The match is impressive in some details but  not perfect. There is an anomalous overall coefficient of $2\csc{\pi \beta \over 2}$, with $\beta$ a sum of external conformal weights as well as a  product of  cosines of the weights appearing in the integral blocks which we do not understand.  Speculations on the origin of these anomalous factors are given in section \ref{ssec:ime}. 

A fascinating feature of this expansion is that the conformal block expansion in massive scalar exchanges pertains even to the pure $\lambda \phi^4$ scattering in which there are no massive scalars, or indeed any intermediate states! This can be traced back to the fact that the Mellin transform from momentum space to celestial amplitudes is well defined only in theories with soft UV behavior. In scale invariant theories like $\lambda \phi^4$ or YM theory the transforms are only marginally convergent and a regulator is needed to define them. In  $\lambda \phi^4$ the regulation effectively amounts to replacing the contact interaction with a massive scalar and taking the infinite mass limit for fixed $g/m$. But since celestial amplitudes have no mass scale the massive scalar does not decouple in this limit! This resonates both with string theory lore and with observations in \cite{Arkani-Hamed:2020gyp} that 
point-like  interactions in quantum gravity should be softened by massive particle exchanges. Of course scalar scattering is not in itself quantum gravity, but the argument that it should  be embedded in  a well-behaved CCFT requires quantum gravity \cite{Kapec:2014opa}. 

Another interesting feature is the CFT$_2$ origin of the $\delta(z-\bar z)$ singularity  in the celestial amplitude, where $z$ is the 2D conformal cross ratio. In momentum space, one starts with a product of four delta functions enforcing momentum conservation. Mellin transforms eliminate three of the four delta functions, with $\delta(z-\bar z)$ remaining as the constraint that all four insertions lie on a celestial equator. In a unitary, rational, local  CFT$_2$,  the 4-point function can never be singular at non-coincident points. However here translation invariance ({i.e.} momentum conservation) shifts $\Delta \to \Delta+1$ and thereby requires an infinite number of primaries.  We see explicitly herein that summing over  these primaries  cancels the finite contribution from the light-ray continuum, while the remaining integral over  light-ray blocks can be shown to produce the delta function.

Our work builds on  several  related previous analyses. Lam and Shao \cite{Lam:2017ofc}  performed a partial wave expansion for scalar 4-point scattering in three dimensions and verified a factorization formula implied by the optical theorem. 
A partial wave expansion of a scalar 4-point in a different channel and a YM 4-point was given by Nandan, Schreiber, Volovich and Zlotnikov in \cite{Nandan:2019jas}. Fan, Fotopoulos, Stieberger, Taylor and Zhu \cite{Fan:2021isc} gave a conformal block decomposition for this 4 point after shadowing an external leg. They also found light-ray-like states (with integral weights) but did not match OPE coefficients. 

This paper is organized as follows. Section \ref{cel-a} reviews some basic celestial formulae and establishes conventions. Section \ref{ss} presents the massless scalar 4-point amplitude mediated by massive scalar exchange  in the conformal primary  basis. In section \ref{pwb} we review completeness properties of partial waves on the unitary principal series. In section \ref{sec:cbd} we transform the celestial amplitude to that basis, deform the contour integral over the unitary principal series and derive the conformal block expansion from pole contributions. Properties of the light transform are discussed and the explicit bulk wavefunction of exchanged states is derived. Section \ref{sec:ca} describes the infinite-mass limit and the resulting contact amplitude. Appendix \ref{cbd} presents two non-trivial checks of our main formula \eqref{res}. The normalization of the massive celestial two-point function is computed in appendix \ref{m2ptf}. Appendix \ref{app:t} provides an interpretation of the overall $\csc{\pi \beta \over 2}$ as arising from a sum over off-shell exchanges. The conformal block decomposition of the full 4-point celestial amplitude of massless scalars mediated by massive exchange is worked out in appendix \ref{mm}.

\section{Preliminaries}
\label{cel-a}

Momentum-space scattering amplitudes can be mapped to correlation functions on the celestial sphere transforming covariantly under the  $SL(2, \mathbb{C})$ Lorentz group \cite{Pasterski:2016qvg,Pasterski:2017kqt}. In this section, we review the simplest case of celestial amplitudes associated with the scattering of 4 massless scalars.  We will treat $z$ and $\bz$ as independent real variables. This amounts to considering scattering in (2,2) signature Klein space where massless momenta can be parametrized by their energy $\omega_i$ and celestial coordinates $(z_i,\bz_i)$ as\footnote{These are related to the Klein space metric $ds^2 = -(dX^0)^2 + (dX^1)^2-(dX^2)^2 + (dX^3)^2$ by $X^{\mu} = \frac{1}{2}\left(u n^{\mu} + r q^{\mu} \right),$ where $n^{\mu} = (1, 0, 0, -1)$.} 
\be 
\label{mp}
p_i^{\mu} = \epsilon_i \omega_i q^{\mu}(z_i, \bz_i), \quad q^{\mu}(z, \bz) =  \Big( 1 + z \bz, z + \bz, z - \bz, 1 - z \bz \Big).
\ee
4-point celestial amplitudes of massless scalars were shown in \cite{Gonzalez:2020tpi,Arkani-Hamed:2020gyp} to be related to their momentum-space counterparts by a Mellin transform,
\be 
\label{ca}
\begin{split}
{\mathcal{A}}(z_i, \bar{z}_i; \Delta_i) =K(z_i, \bar{z}_i) X(z, \beta) \int_0^{\infty} d\omega \omega^{\beta - 1}\mathcal{M}(\omega^2, -z \omega^2).
\end{split}
\ee
Here
\be
\label{ccp}
K(z_i, \bz_i) \equiv \prod\limits_{i < j} z_{ij}^{\frac{h}{3} - h_i - h_j} \bar{z}_{ij}^{\frac{\bar{h}}{3} - \bar{h}_i - \bar{h}_j}
\ee
is a (symmetric) conformally covariant factor and
\be 
X(z, \beta) \equiv  2^{-\beta - 2} |z(1 - z)|^{\frac{1}{6}(\beta + 4)} \delta(z - \bz).
\ee
$\mathcal{M}(s, t)$ is the 4-point momentum-space scattering amplitude of scalars upon stripping off the momentum conserving delta function,  $h_i = \bar{h}_i = \frac{\Delta_i}{2}$ are conformal weights,
\be 
\beta = \sum_{i = 1}^4 \Delta_i - 4
\ee
and $z$ is the standard cross ratio
\be \label{sx}
z = \frac{z_{13} z_{24}}{z_{12} z_{34}}.
\ee

\begin{figure}[H]
 \centering


\tikzset{
pattern size/.store in=\mcSize, 
pattern size = 5pt,
pattern thickness/.store in=\mcThickness, 
pattern thickness = 0.3pt,
pattern radius/.store in=\mcRadius, 
pattern radius = 1pt}
\makeatletter
\pgfutil@ifundefined{pgf@pattern@name@_7ezmtp7sj}{
\pgfdeclarepatternformonly[\mcThickness,\mcSize]{_7ezmtp7sj}
{\pgfqpoint{0pt}{0pt}}
{\pgfpoint{\mcSize+\mcThickness}{\mcSize+\mcThickness}}
{\pgfpoint{\mcSize}{\mcSize}}
{
\pgfsetcolor{\tikz@pattern@color}
\pgfsetlinewidth{\mcThickness}
\pgfpathmoveto{\pgfqpoint{0pt}{0pt}}
\pgfpathlineto{\pgfpoint{\mcSize+\mcThickness}{\mcSize+\mcThickness}}
\pgfusepath{stroke}
}}
\makeatother
\tikzset{every picture/.style={line width=0.75pt}} 

\begin{tikzpicture}[x=0.75pt,y=0.75pt,yscale=-1,xscale=1]

\draw  [pattern=_7ezmtp7sj,pattern size=6pt,pattern thickness=0.75pt,pattern radius=0pt, pattern color={rgb, 255:red, 0; green, 0; blue, 0}] (270,150) .. controls (270,127.91) and (287.91,110) .. (310,110) .. controls (332.09,110) and (350,127.91) .. (350,150) .. controls (350,172.09) and (332.09,190) .. (310,190) .. controls (287.91,190) and (270,172.09) .. (270,150) -- cycle ;
\draw    (241,82) -- (281,122) ;
\draw    (339,177) -- (379,217) ;
\draw    (338,122) -- (378,82) ;
\draw    (242,219) -- (282,179) ;
\draw    (190,150) -- (248,150) ;
\draw [shift={(250,150)}, rotate = 180] [color={rgb, 255:red, 0; green, 0; blue, 0 }  ][line width=0.75]    (10.93,-3.29) .. controls (6.95,-1.4) and (3.31,-0.3) .. (0,0) .. controls (3.31,0.3) and (6.95,1.4) .. (10.93,3.29)   ;
\draw    (309.39,262.11) -- (309.98,213) ;
\draw [shift={(310,211)}, rotate = 450.68] [color={rgb, 255:red, 0; green, 0; blue, 0 }  ][line width=0.75]    (10.93,-3.29) .. controls (6.95,-1.4) and (3.31,-0.3) .. (0,0) .. controls (3.31,0.3) and (6.95,1.4) .. (10.93,3.29)   ;
\draw    (242,219) -- (260.59,200.41) ;
\draw [shift={(262,199)}, rotate = 495] [color={rgb, 255:red, 0; green, 0; blue, 0 }  ][line width=0.75]    (10.93,-3.29) .. controls (6.95,-1.4) and (3.31,-0.3) .. (0,0) .. controls (3.31,0.3) and (6.95,1.4) .. (10.93,3.29)   ;
\draw    (379,217) -- (360.41,198.41) ;
\draw [shift={(359,197)}, rotate = 405] [color={rgb, 255:red, 0; green, 0; blue, 0 }  ][line width=0.75]    (10.93,-3.29) .. controls (6.95,-1.4) and (3.31,-0.3) .. (0,0) .. controls (3.31,0.3) and (6.95,1.4) .. (10.93,3.29)   ;
\draw    (281,122) -- (262.41,103.41) ;
\draw [shift={(261,102)}, rotate = 405] [color={rgb, 255:red, 0; green, 0; blue, 0 }  ][line width=0.75]    (10.93,-3.29) .. controls (6.95,-1.4) and (3.31,-0.3) .. (0,0) .. controls (3.31,0.3) and (6.95,1.4) .. (10.93,3.29)   ;
\draw    (338,122) -- (356.59,103.41) ;
\draw [shift={(358,102)}, rotate = 495] [color={rgb, 255:red, 0; green, 0; blue, 0 }  ][line width=0.75]    (10.93,-3.29) .. controls (6.95,-1.4) and (3.31,-0.3) .. (0,0) .. controls (3.31,0.3) and (6.95,1.4) .. (10.93,3.29)   ;

\draw (230,219) node [anchor=north west][inner sep=0.75pt]    {$1$};
\draw (379,217) node [anchor=north west][inner sep=0.75pt]    {$2$};
\draw (230,66) node [anchor=north west][inner sep=0.75pt]    {$3$};
\draw (376,66) node [anchor=north west][inner sep=0.75pt]    {$4$};
\draw (207,129) node [anchor=north west][inner sep=0.75pt]    {$t$};
\draw (315,233) node [anchor=north west][inner sep=0.75pt]    {$s$};

\end{tikzpicture}
 \caption{4-point kinematics with particles $1,2$ incoming and particles $3,4$ outgoing. We will study the $13-24$ OPE limit ($z \rightarrow 0$) in the CCFT which corresponds to the collinear limit of $1$ and $3$ in the bulk.}
 \label{fig:1}
  \end{figure}
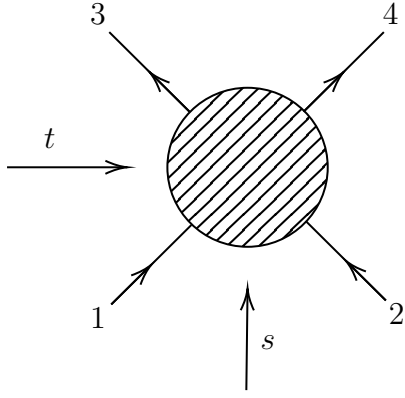

In  terms of the 
 Mandelstam invariants 
\be 
s = -(p_1 + p_2)^2, \quad t = -(p_1 + p_3)^2,
\ee
one finds 
 the cross-ratio is 
 \be z \equiv -\dfrac{t}{s}.\ee In this paper we will study the physical configuration $s \geq 0, ~ t\in  [-s,0] \implies z \in [0,1]$, corresponding to particles 1 and 2 incoming ($\epsilon_1 = \epsilon_2 = -1$) and particles 3 and 4 outgoing ($\epsilon_3 = \epsilon_4 = 1$). We illustrate this in figure \ref{fig:1}.

While \eqref{ca} resembles a 2D CFT 4-point function constrained by global conformal symmetry, it enjoys two additional properties: upon stripping off the conformally covariant prefactor \eqref{ccp}, it depends on the external weights only through their sum $\beta$ and it is multiplied by a delta function enforcing that the 4 points lie on a circle. Both these properties follow more generally from translation invariance \cite{Law:2019glh}.

In this paper we will be interested in the expansion of celestial amplitudes \eqref{ca} around the $z \rightarrow 0$ limit which should encode information about the operator product expansions (OPE) of particles 1, 3 and 2, 4. To this end, it will be convenient to relate $K(z_i, \bz_i)$ to the  conformally covariant structure symmetric under 1-3 and 2-4 exchanges,
\be 
I_{13-24}(z_i, \bz_i) \equiv \dfrac{\left(\dfrac{z_{34}}{z_{14}} \right)^{h_{13}} \left(\dfrac{z_{14}}{z_{12}} \right)^{h_{24}}}{z_{13}^{h_1 + h_3} z_{24}^{h_2 + h_4}} \dfrac{\left(\dfrac{\bz_{34}}{\bz_{14}} \right)^{\bh_{13}} \left(\dfrac{\bz_{14}}{\bz_{12}} \right)^{\bh_{24}}}{\bz_{13}^{\bh_1 + \bh_3} \bz_{24}^{\bh_2 + \bh_4}}.
\ee
We find
\be 
\label{K-I}
\begin{split}
K(z_i, \bar{z}_i) X(z, \beta) &= \prod\limits_{i < j} z_{ij}^{\frac{h}{3} - h_i - h_j} \bar{z}_{ij}^{\frac{\bar{h}}{3} - \bar{h}_i - \bar{h}_j}  2^{-\beta - 2} |z(1 - z)|^{\frac{1}{6}(\beta + 4)} \delta(z - \bar{z}) \\
&= I_{13-24}(z_i, \bar{z}_i)   2^{-\beta - 2} |z|^{2 + \beta/2} |1 - z|^{h_{13} - h_{24}} \delta(z - \bar{z}).
\end{split}
\ee
This allows us to put generic celestial 4-point amplitudes into the form
\be 
\label{m-exch-am}
\mathcal{A}(z_i, \bz_i; \Delta_i) = I_{13-24}(z_i, \bar z_i)\, {f}(z, \bz),
\ee
with ${f}(z,\bz)$ obtained from \eqref{K-I} and \eqref{ca}. 

In CCFT we are a priori interested in arbitrary complex dimensions $(h_i,\bar h_i)$ for which Mellin integrals or  other quantities may be  ill-defined. Wherever such ambiguities arise  we invoke  analytic continuation from real values of $(h_i,\bar h_i)$. A discussion of the analytic properties of 4-point scattering can be found in \cite{Arkani-Hamed:2020gyp}.

\section{Scalar scattering}
\label{ss}
\subsection{Massive exchange}

Here we consider the
scattering of 4 massless scalars $\phi$ mediated by a scalar $\Psi$ of mass $m$. The stripped momentum-space amplitude is 
\be 
\label{sa}
\mathcal{M}(s, t) = -g^2 \Big(\frac{1}{s - m^2} + \frac{1}{t - m^2} + \frac{1}{-t - s - m^2} \Big).
\ee
Plugging \eqref{sa} into \eqref{ca} and evaluating the integral using contours, we find the celestial amplitude takes the form \eqref{m-exch-am} with 
\be 
\begin{split}
\label{4-sc-sc}
{f}_{stu}(z, \bz) &= N_{gm}(\beta) \delta(z - \bz) |z|^{2}|z - 1|^{h_{13} - h_{24}} \left[ e^{i \pi\beta/2} |z|^{\beta/2} + 1 + \left|\frac{z}{z - 1} \right|^{\beta/2}\right].
\end{split}
\ee 
The $z$-independent prefactor is 
\be \cb=\frac{g^2 \pi}{8 m^2} \frac{\left(m/2\right)^{\beta}}{\sin \pi \beta/2},\ee
with $g$ the strength of the $\phi^2 \Psi$ interaction and $h_{ij} \equiv h_i - h_j.$ This agrees with \cite{Nandan:2019jas} upon redefinition of the cross-ratio. 

It is possible to consistently project out the $s$, $t$ or $u$ channel exchanges. In a theory with two massless scalars $\phi_1, \phi_2$ which couple to $\Psi$ via equal strength interactions $g \phi_i^2 \Psi$, the $\phi_1 \phi_2 \rightarrow \phi_1 \phi_2$ amplitude contains only the second  $t-$channel term and ${f}_{stu}(z, \bz)$ simply reduces to 
\be 
\label{fz}
{f}_t(z, \bz) = N_{gm}(\beta) |z|^{2}|z - 1|^{h_{13} - h_{24}}\delta(z - \bz) .
\ee
For simplicity we concentrate on this 4-point  scattering in the following. However we note that all three terms in \eqref{4-sc-sc} have an albeit more complicated conformal block decomposition in the $13-24$ channel (this case is deferred to the appendix).  While discrete and continuous exchanges are still present, with discrete states of dimensions differing by integers, these two kinds of contributions are in this case not obviously related by a light-ray transform, their dimensions are in general $\beta$ dependent and $J \neq 0$ contributions are allowed. Moreover, the coefficients of the decompositions don't factorize, although simplifications may occur at special values of the external dimensions. We leave a full analysis to future work.

\subsection{Contact interaction}\label{subsec:contact}

It is not hard to see from a change of variables in \eqref{ca} that any $s$-independent amplitude $\mathcal{M}$ will take the form \eqref{m-exch-am} with ${f}(z, \bz)$ having the same $z$-dependence as \eqref{fz}. For example 
tree-level $\phi^4$ 4-point contact scattering has
\be 
\label{msa}
\mathcal{M}(s, t) = \lambda.
\ee
Plugging this into \eqref{ca} one immediately finds 
\be 
\label{contact-am}
\begin{split}
{f}_\lambda(z, \bz)
&= \frac{\pi \lambda}{2} \delta(\beta )  |z|^2 |z - 1|^{h_{13} - h_{24}} \delta(z - \bz).
\end{split}
\ee

In fact one can show that \eqref{4-sc-sc} reduces to the 4-point contact interaction as $m \rightarrow \infty$ for fixed $g/m$. The following identity which holds for purely imaginary $\beta$ or equivalently external states on the principal series will be useful\footnote{A prescription for analytically continuing this result to $\beta \in \mathbb{C}$ was proposed in \cite{Donnay:2020guq}.}
\be
\label{id}
    \lim_{z \to 0} z^{-\beta} = \frac{1}{\Gamma(\beta)} \lim_{z\to 0} \int_0^\infty d \omega \, \omega^{\beta-1} e^{-\omega z} = \frac{1}{\Gamma(\beta)} \int_0^{\infty} d\omega\, \omega^{\beta - 1} = 2 \pi \beta \delta(\beta).
\ee
Looking now at the prefactor of \eqref{4-sc-sc}, we see
\be \label{inf-m} \lim_{m \rightarrow \infty}\cb = \frac{g^2}{8 m^2} \frac{ \pi \beta}{\sin \pi \beta/2} 2 \pi \delta( \beta) = \frac{g^2 \pi}{2 m^2} \delta(\beta).
\ee
Since $\beta$ is set to 0, we deduce that the $|z|^{\beta/2}$ and $\left|\frac{z}{z - 1}\right|^{\beta/2}$ terms are set to 1 and the celestial amplitude collapses to
\be
\label{cont}
  \lim_{m \rightarrow \infty} f_{stu}(z, \bz) =  \frac{3 g^2 \pi}{2m^2}  \delta(\beta) |z|^2 |1-z|^{h_{13} - h_{24}} \delta(z-\bar z).
\ee
This is precisely the amplitude of the 4-point contact interaction \eqref{contact-am} subject to the identification
\be 
\frac{3 g^2}{m^2} = \lambda.
\ee

\section{Partial wave basis}
\label{pwb}

Any conformally covariant 4-point function admits a conformal partial wave expansion in a complete  basis of orthonormal solutions to the two-particle conformal Casimir equation. One then hopes to reexpress this decomposition as a sum or integral over conformal blocks associated to intermediate conformal primary states. This representation of the four-point function gives information  
both about the complete basis of primary states and their OPEs. In particular when a complete basis is identified, their OPE should close and no other primaries  should appear in the conformal block expansion. In CCFT one typically discusses conformal primary wave functions as well as their shadows and light transforms with arbitrary complex weight, with soft currents and Goldstone bosons arising at real integer weights. All of these together are clearly an overcomplete basis. Which if any subset comprise a complete basis on which all the symmetries act is an open question which we hope the current work will help address. 

In a standard 2D CFT with a non-degenerate spectrum, the OPE 
\be 
\label{OPE}
\mathcal{O}_1(z_1, \bz_1) \mathcal{O}_3(z_3, \bz_3) \sim \sum_{\mathcal{O}_k~{\rm primary}} z_{13}^{h_k - h_1 - h_3} \bz_{13}^{\bh_k - \bh_1 - \bh_3} C_{13}^{\ \ k} \mathcal{O}_k(z_3, \bz_3) + \cdots ,
\ee
where ... includes a sum over descendants, allows CFT 4-point correlation functions to be written as \cite{DiFrancesco:1997nk} 
\be
\label{4pt-general}
\Big< \prod_{i=1}^4 \mathcal{O}_i^{h_i, \bh_i}(z_i, \bz_i) 
 \Big> = I_{13-24}(z_i, \bz_i) \sum_{\mathcal O_k \text{ primary}} C_{13k} C_{24}^{\ \ k}\, k^{\{h_i, \bh_i\}}_{h_{k}, \bh_{k}} (z, \bar z).
 \ee
Here $k_{h, \bh}^{\{h_i, \bh_i\}}$ are $\SL(2, \CC)$ conformal blocks which resum the contribution from a primary $\mathcal O_{h,\bh}$ and its SL$(2,\mathbb{C})$ descendants appearing in the $\mathcal{O}_1, \mathcal{O}_3$ OPE and $z = \frac{z_{13}z_{24}}{z_{12}z_{34}}$ as in \eqref{sx}. From now on, we will suppress any labels indicating the dependence of the conformal blocks on external dimensions. $C_{ijk}$ are coefficients of three-point functions
  \be
\label{3pt-general}
    \left<\O_i(z_i, \bz_i) \O_j(z_j, \bz_j) \O_k(z_k, \bz_k) \right> = \frac{C_{ijk}}{z_{ij}^{h_i + h_j - h_k} z_{jk}^{h_j +h_k - h_i} z_{ik}^{h_i + h_k - h_j} \bz_{ij}^{\bh_i + \bh_j - \bh_k} \bz_{jk}^{\bh_j+\bh_k - \bh_i} \bz_{ik}^{\bh_i + \bh_k - \bh_j} } 
\ee
and $C_{ij}^{\ k} = C_{ijl} D^{lk}$ where 
\be
\label{2pt-general}
    \left<\O_i(z_i, \bz_i) \O_j(z_j, \bz_j) \right> = \frac{D_{ij}}{z_{ij}^{h_i +h_j} \bz_{ij}^{\bh_i + \bh_j}}.
\ee
Conformal invariance requires $D_{ij} = 0$ for operators of different scaling dimension.
\eqref{4pt-general} can be derived by replacing $\mathcal{O}_1 \mathcal{O}_3$ on the LHS by the OPE \eqref{OPE} (including descendants) and resuming the three-point functions \eqref{3pt-general} and their analogs involving descendants of one of the operators into $k_{h, \bar{h}}(z, \bz).$ 

Alternatively, conformal symmetry implies that the blocks are eigenfunctions of the conformal Casimir acting on operators 1 and 3. This reduces to the following differential equation
\be 
\label{CCas}
\left(\mathcal{D}_z + \mathcal{D}_{\bz} \right) k_{h, \bh}(z, \bz) = \left[h (h - 1) + \bh(\bh - 1)\right] k_{h, \bh}(z, \bz),
\ee
where 
\be 
\label{dz}
\mathcal{D}_z = z^2(1 - z) \frac{\p^2}{\p z^2} - (1   - h_{13} + h_{24}) z^2 \frac{\p }{\p z}  + h_{13} h_{24} z, ~~ h_{ij} \equiv h_{i} - h_j. 
\ee
It turns out that \eqref{CCas} admits a complete, orthogonal basis of solutions \cite{Dolan:2011dv,Karateev:2018oml,Rutter:2020vpw}. We treat $z, \bz$ as real independent variables, or equivalently we work in $(2,2)$ Klein space. In this  case
\be 
\label{psi2}
\Psi_{h, \bar{h}}(z, \bz) = \Psi_h(z) \Psi_{\bar{h}}(\bz)
\ee
forms such a set provided $\Psi_{h}$ and $\Psi_{\bar{h}}$ are both complete and orthogonal solutions to the SL$(2, \mathbb{R})$ conformal Casimir equations
\be
\label{1dCCas}
	\mathcal{D}_z \Psi_h(z) = h (h-1) \Psi_h (z), \quad 	\mathcal{D}_{\bz} \Psi_{\bar{h}}(\bz) = \bh (\bh-1) \Psi_{\bh} (\bz).
\ee 
Symmetry of \eqref{1dCCas} under $h \rightarrow 1 -  h$ implies $\Psi_h$ is a linear combination of the $\SL(2, \RR)$ conformal blocks
\be
\begin{split}
\label{1dcb}
k_h(z) = z^h \pFq{2}{1}{h - h_{13}, h + h_{24}}{2h}{z}
\end{split}
\ee
and their shadows $k_{1 - h}(z)$. 
For real external dimensions, it was shown in \cite{Hogervorst:2017sfd,Rutter:2019hqc} that $\mathcal{D}_z$ is self-adjoint with respect to the inner product
\be 
\label{inpr}
\langle F, G \rangle = \int_0^1 \frac{dz}{z^2} (1-z)^{-h_{13} + h_{24}} F^*(z) G(z).
\ee

A generic set of $\mathcal{D}_z$ eigenfunctions with real eigenvalues $\alpha^2 - \frac14$ and $h = \frac{1}{2} + \alpha$, $\alpha \in i \mathbb{R}$  is
\be\label{1DCCasSoln}
    \Psi_{\alpha}(z) = A k_{1/2 + \alpha}(z) + B k_{1/2 - \alpha}(z), \quad \alpha \in i \RR.
\ee
Imposing the boundary condition $\Psi(1)=1$ fixes $A$ and $B$ and yields a basis of solutions \cite{Rutter:2019hqc}
\be 
\label{1DCPW}
\begin{aligned}
	\Psi_\alpha &= \frac{1}{2}\Big(Q(\alpha) k_{1/2 + \alpha}(z) + Q(- \alpha) k_{1/2 - \alpha}(z) \Big),\\
	Q(\alpha) &= \frac{2 \Gamma(-2\alpha) \Gamma(1 - h_{13} + h_{24})}{\Gamma(\frac{1}{2} - \alpha - h_{13}) \Gamma(\frac{1}{2} - \alpha + h_{24})}.
\end{aligned}
\ee
The conformal partial waves \eqref{1DCPW} are real-valued ($\Psi^*_\alpha  = \Psi_{-\alpha} = \Psi_{\alpha}$) for real $h_{13}, h_{24}$, orthogonal with respect to the inner product \eqref{inpr}
\be 
\label{ip}
	\braket{\Psi_\alpha, \Psi_{\alpha'}} = \int_0^1 \frac{dz}{z^2} (1-z)^{-h_{13}+h_{24}} \Psi_{\alpha}(z)\Psi_{\alpha'}(z) = 2 \pi i \frac{N(\alpha)}{2} [\delta(\alpha-\alpha') + \delta(\alpha + \alpha')],
\ee
with 
\be
	\quad N(\alpha) = \frac{Q(\alpha) Q(-\alpha)}{2}
\ee
and obey the completeness relation
\be 
\label{completeness}
\frac1{2\pi i}\int_{-i\infty}^{i\infty} \frac{d\alpha}{N(\alpha)} \Psi_\alpha (z) \Psi_{\alpha}(w) = z^2 (1-z)^{h_{13}-h_{24}} \delta(z-w). 
\ee
Using a hypergeometric identity  $\Psi_\alpha$ can be compactly written as
\be 
\label{psic}
\begin{split}
\Psi_{\alpha}(z) = z^{h_{13}} \pFq{2}{1}{\frac12 + \alpha - h_{13} , \frac12 - \alpha -h_{13}}{1 - h_{13} + h_{24}}{-\frac{1-z}{z}}.
\end{split}
\ee

 This argument can be easily generalized  \cite{Rutter:2020vpw} to show that 
\be 
\label{cpw-a}
\Psi_{\alpha, \bar{\alpha}}(z, \bz) = \Psi_{\alpha}(z) \Psi_{\bar{\alpha}}(\bz), \quad \alpha, \bar{\alpha} \in i \mathbb{R}
\ee
form a basis of the $SL(2, \mathbb{R}) \times SL(2, \mathbb{R})$ conformal Casimir equation on the Lorentzian square $z, \bz \in [0,1]$. It follows from \eqref{ip} and \eqref{completeness} that any conformally invariant function $f(z, \bz)$ can be decomposed as 
 \be\label{2Dainv}
	f(z, \bz) = \frac{1}{2\pi i} \int_{-i\infty}^{i\infty} \frac{d \alpha}{N(\alpha)} \frac{1}{2\pi i}\int_{-i\infty}^{i\infty} \frac{d\ba}{N(\ba)} \hat f(\alpha, \ba) \Psi_{\alpha, \ba}(z, \bz),
\ee
where 
\be\label{2Datrans}
	\hat f(\alpha, \bar{\alpha}) = \int_0^1 \frac{dz}{z^2} (1-z)^{-h_{13}+h_{24}} \int_0^1 \frac{d\bz}{\bz^2} (1-\bz)^{-h_{13}+h_{24}} f(z, \bz)  \Psi_{\alpha, \ba}(z, \bz).
\ee
We emphasize that \eqref{cpw-a} form a basis on the Lorentzian square $z, \bz \in [0,1]$ while standard $SL(2, \mathbb{C})$ conformal partial waves such as in \cite{Dolan:2011dv} form a basis on the Euclidean plane. It is this property that simplifies the analysis of celestial 4-point amplitudes with non-trivial support for $z \in [0,1]$ and $z = \bz$ which we turn to next.

\section{Conformal block decomposition}
\label{sec:cbd}

In this section we decompose the tree-level 4-scalar celestial amplitude  in the partial wave  basis \eqref{psi2}, and then use contour deformation to reexpress it as a sum plus integral over exchanges of conformal blocks associated to primary operators multiplied by the appropriate squared OPE coefficients.

We take $f(z,\bz)$ to be the t-channel amplitude $f_t(z,\bz)$ in   \eqref{fz}. We will first take the external dimensions to be real, and then obtain the final result for external dimensions on the principal series by analytic continuation. We begin by evaluating \eqref{2Datrans} for $f={f}_t$ 
\be 
\label{inv1}
\begin{aligned}
\hat f_t(\alpha, \ba) &= \cb\int_0^1 \frac{dz}{z^2} (1-z)^{-h_{13}+h_{24}} \int_0^{1} \frac{d\bz}{\bz^2}  (1-\bz)^{-\bh_{13}+\bh_{24}}\\
&\times z^2 (1 - z)^{h_{13} - h_{24}} \delta(z - \bz) \Psi_{\alpha}(z) \Psi_{\ba}(\bz)\\
&= \cb\int_0^1 \frac{dz}{z^2}  (1 - z)^{ -h_{13} + h_{24}} \Psi_{\alpha}(z) \Psi_{\ba}(z),
\end{aligned}
\ee
where 
we used  $h_{13} = \bh_{13}, h_{24} = \bh_{24}$ for the four-point scalar amplitude. We recognize the last line of \eqref{inv1} as the orthogonality relation \eqref{ip}, yielding
\be\label{inv2}
	\hat f_t(\alpha, \ba) =  N_{gm}(\beta) \pi i \frac{Q(\alpha) Q(-\alpha)}{2} \left[ \delta(\alpha - \ba) + \delta(\alpha + \ba)\right].
\ee
Plugging \eqref{inv2} back into \eqref{2Dainv} results in the partial wave decomposition
\be
    z^2 (1-z)^{-h_{13}+h_{24}} \delta(z-\bz) = \frac{1}{2\pi i} \int_{-i\infty}^{i\infty} \frac{d\alpha}{N(\alpha)} \Psi_\alpha(z) \Psi_\alpha(\bar z).
\ee
Next we use \eqref{1DCPW} and deform the contour in the right-hand complex $\alpha$ plane for the term proportional to  $k_{\frac{1}{2} + \alpha}(z) k_{\frac{1}{2} + \alpha}(\bz)$. This can be evaluated at the poles of the $Q(\alpha)$ coefficients in \eqref{1DCPW}, yielding 
\be 
\label{res}
\begin{aligned}
f_t(z,\bz)
= &2 \csc{ \pi \beta \over 2}\Big[\
  \sum_{n = 1}^{\infty}  D^{nn}  C_{13n} C_{24n}  \cos\pi \!\left(\tfrac{n}{2} + h_{13}\right) \cos\pi \!\left(\tfrac{n}{2} + h_{24}\right) k_{\frac{1+n}{2}}(z) k_{\frac{1+n}{2}}(\bz)\cr&~~~~~~~~~~~~~ +  \frac{1}{2}\int_{-i\infty}^{i\infty} d\alpha C_{13\alpha}^L C_{24\alpha}^L D^{L,\alpha \alpha} k_{\frac{1}{2} + \alpha}(z) k_{\frac{1}{2} - \alpha}(\bz)\Big],
\end{aligned}
\ee
\be 
\label{OPEs}
\begin{split}
C_{ijn} &= \frac{g}{m^4}\left({m\over 2}\right)^{2h_i+2h_j}\mathrm{B}\left(\frac{1+n}{2} + h_{ij}, \frac{1+n}{2} - h_{ij}\right),\\
C_{ij\alpha}^L &= -\pi i\frac{g}{m^4}\left({m\over 2}\right)^{2h_i+2h_j}\frac{1}{\alpha}
\end{split}
\ee
and 
\be 
D^{nn}  = \frac{n m^2}{2\pi}, \quad D^{L, \alpha \alpha} = -\frac{m^2\alpha^2}{\pi^2 i}.
\ee
We demonstrate explicitly in appendix \ref{cbd} that the right-hand-side of \eqref{res} vanishes for $z \neq \bz$ due to a cancellation between the two terms.  Here $D^{nn}$ and $D^{L, \alpha \alpha}$ are the (inverse of the) coefficient of the celestial two-point function of massive scalars and its light transform respectively. These are derived in appendix \ref{m2ptf} and section \ref{Lro}.

\subsection{Integer  mode exchanges}
\label{ssec:ime}
Consider the first term in \eqref{res}. The coefficients \eqref{OPEs} are precisely the previously  known OPEs of two massless and one massive scalar, as computed from  the three-point celestial amplitude of two massless and one massive particle\cite{Lam:2017ofc}\footnote{In (3,1) signature the three point functions of one  massless incoming, one massless  outgoing  and one massive incoming or outgoing  particle vanishes. On the other hand, in the (2,2) case considered here, this kinematic configuration is allowed.}. Hence this term corresponds  to  a tower of massive exchanges of positive integer dimensions $\Delta = n + 1$ and spin $J = 0.$
The factor of $n$ in \eqref{res} arises from the normalization of the celestial two-point function of massive scalars of dimension $\Delta = 1 + n$ as detailed in appendix \ref{m2ptf}. 

This remarkable match still leaves unexplained the prefactor of $2\csc{\pi\beta \over 2}$ as well as the product of $ \cos\pi({n\over 2}+h_{ij})$. The first   factor, which depends only on $\beta$ arises in a  universal manner  (see appendix \ref{app:t}), and has analogs present in all previous discussions of conformal factorization \cite{Lam:2017ofc,Nandan:2019jas,Fan:2021isc}.  Interestingly it drops out of  the residues of the poles of the amplitude (indeed it is the source of them), which were argued in \cite{Arkani-Hamed:2020gyp} to contain all the information about the low energy effective action.  The form of the second factor suggests it might be absorbed in the OPE coefficients. Moreover, it disappears when restricted to integer weights.  Finally it is interesting to note that similar factors are generated in Lorentzian CFT correlators involving terms with different time orderings (see for example section 2.5 of \cite{Kravchuk:2018htv}). For now we simply leave these discrepancies to hopefully be resolved in future work. 

The second term in \eqref{res} corresponds to a continuum exchange  of operators of $\Delta = 1$ and $J = -i \lambda$ which coincide with dimensions of operators obtained from scalars on the principal series (ie. with $\Delta = 1 + i\lambda$) via the so-called light-ray transform \cite{Kravchuk:2018htv} which takes $(\Delta, J) \rightarrow (1- J, 1 - \Delta)$. The next section  considers these in detail.

\subsection{Light transform exchanges}
The conformal block decomposition \eqref{res} contains terms involving exchanges of dimension $\Delta = 1$ and imaginary spin $J=h-\bar h$. Such contributions are not allowed in Euclidean CFT, where the solutions to \eqref{CCas} are constrained to be single-valued and $J$ is accordingly quantized. However we are working in $(2,2)$ Klein space and the boundary supports a Lorentzian $(1,1)$ CFT. There is no quantization of spin in $(1,1)$ signature and indeed imaginary spin values are regularly encountered  in conformal block decompositions \cite{Caron-Huot:2017vep,Kravchuk:2018htv,Kologlu:2019mfz}. In this section we show that the imaginary spin exchanges appearing in the decomposition of Lorentzian celestial 4-point amplitudes correspond to new Klein space solutions of the massless and massive wave equations which have not previously been discussed.\footnote{Light transforms also exist in $(3,1)$ for integer $J$.} The light transform relates these  to scalar solutions on the principal series $\Delta=1+i \lambda, J = 0$. Moreover, we demonstrate that the coefficient of the imaginary spin block is proportional  to the squared OPE  coefficients  of two massless scalars and a light-transformed massive scalar. 

\subsubsection{Boundary}

In a generic conformal field theory, the light transform is an integral of a conformal primary operator $\mathcal O$ along a light ray. In $1+1$ dimensions, there are two null directions, corresponding to $z =0$ or $\bar z = 0$. Therefore, light transforms of spin $0$ operators correspond to the holomorphic or anti-holomorphic integrals given by\footnote{\eqref{light} is related to the formula given in \cite{Kravchuk:2018htv} by a change of variables.}
	\begin{equation}
		\label{light}
	    \begin{split}
		\mathbf L[\Phi_\Delta] (z, \bar z, +)&=\int_{-\infty}^{\infty} dz' \frac{1}{(z' - z)^{2-\Delta}} \Phi_{\Delta}(z', \bar z), \\
		\mathbf L[\Phi_\Delta] (z, \bar z, -)&=\int_{-\infty}^{\infty} d\bar z' \frac{1}{(\bar z' - \bar z)^{2-\Delta}} \Phi_{\Delta}(z, \bar z').
	\end{split}
	\end{equation}
Notice that the product of these two  light-ray transforms is  the more commonly encountered shadow transform in Euclidean signature \cite{SimmonsDuffin:2012uy}.  \subsubsection{Bulk}
In this subsection we construct the holomorphic light transforms of bulk scalar primary wave functions as solutions to the scalar wave equation of weights $h' = 1-h$ and $\bar h' = \bar h$ respectively. This corresponds to dimension $\Delta' = 1$ and spin $J' = 1-\Delta$. In particular, light-transformed primaries of operators with dimension $\Delta = 1 + i \lambda$ on the principal series have imaginary spin $J'=-i\lambda$.  

 Massless scalar particles in Klein space  satisfy the wave equation
	\begin{equation}
		\Box \phi = 0, \quad ds^2 = -(dX^0)^2 + (dX^1)^2 -(dX^2)^2 + (dX^3)^2.
	\end{equation}
	With the parameterization of $q$ in \eqref{mp},
	\begin{align}
		q \cdot X &= -(X^0 - X^3) + z (X^1 - X^2) + \bar z (X^1 + X^2) - z \bar z (X^0 + X^3).
	\end{align}
	Massless scalar primaries  take the form\footnote{We suppress the $\pm i\epsilon$ prescription. In Klein space 
	the $\pm$ sign can be absorbed by a rotation of $q$ up to normalization.}\cite{Atanasov:2021oyu}
	\begin{equation}\label{eq:original}
		\phi_{\Delta}(z, \bar z) = \frac{i^{\Delta} \Gamma(\Delta)}{(-q \cdot X)^\Delta}.
	\end{equation}
The holomorphic light-transform of \eqref{eq:original} is 
	\begin{equation}
		\begin{aligned}
		&\int_{-\infty}^{\infty} \frac{dz'}{(z' - z)^{2 - \Delta}} \frac{e^{i \pi \Delta/2}\, \Gamma(\Delta)}{(-q(z', \bar z) \cdot X )^\Delta}\\& = \int_{-\infty}^{\infty} \frac{dz'}{(z' - z)^{2 - \Delta}}\frac{e^{ i \pi \Delta/2}\, \Gamma(\Delta)}{((X^0 - X^3) - \bar z (X^1 + X^2) - z' (X^1 - X^2 - \bar z (X^0 + X^3)) )^\Delta}.
		\end{aligned}
	\end{equation}
		Upon changing variables, this integral becomes proportional to
\be 
\begin{split}
\int_{-\infty}^{\infty} \frac{dz'}{(z' - z)^{2 - \Delta}} \frac{1}{(-q(z',\bz)\cdot X)^{\Delta}} &= \int_{-\infty}^{\infty} dz' z'^{\Delta - 2} (-q\cdot X - z' \p_z q\cdot X)^{-\Delta}\\
& = \frac{(\p_z q\cdot X)^{1 - \Delta}}{-q\cdot X}\int_{-\infty}^{\infty} dz' z'^{\Delta-2}(1 - z')^{-\Delta}.
\end{split}
\ee
The remaining integral is directly related to the  Euler beta function\footnote{This requires a careful choice of branch for the indefinite integral. There exists a second solution $\widetilde{\phi} \propto \frac{(\p_z q\cdot X)^{2 - \Delta}}{q\cdot X}\delta(\p_z q\cdot X)$ with the correct transformation properties whose relevance we leave  to future work.}
\be 
\lim_{\epsilon \rightarrow 0}\int_{-\infty}^{\infty} dz z^{\Delta - 2 + \epsilon} (1 - z)^{-\Delta} = \lim_{\epsilon \rightarrow 0}\frac{-2 \pi^2 i}{\sin \pi \epsilon \Gamma(\Delta - 2 + \epsilon) \Gamma(-\Delta) \Gamma(\epsilon)} = -2\pi i \frac{1}{\Gamma(-\Delta) \Gamma(\Delta - 2)}.
\ee
Putting eveything together, we get the light-transformed field
\be 
\label{lrs}
	\widetilde{\phi}(z, \bar z; X) = 2\pi i e^{i\pi \Delta/2}\frac{(\Delta -1)(\Delta-2)}{\Gamma(-\Delta)}\frac{(\p_z q\cdot X)^{1 - \Delta}}{q\cdot X}.
\ee
\eqref{lrs} has the correct scaling properties, namely under $\bar z \to \alpha \bar z$, $q \cdot X \to \alpha^{1/2} q \cdot X$ and \eqref{lrs} is rescaled by $\alpha^{-1/2 -  i \lambda/2}$. Similarly, $z \to \alpha z$ gives a net factor of $\alpha^{-1/2 + i \lambda/2}$. It follows that \eqref{lrs} have dimension and spin $\Delta = 1, J = -i  \lambda$ consistent with the light-transform of a conformal scalar primary of dimension $\Delta = 1 + i \lambda$.

This analysis is easily generalized to massive primaries \cite{Pasterski:2016qvg} 
	\begin{equation}
		\frac{4\pi}{i m} \frac{(\sqrt{-X^2})^{\Delta- 1}}{(-q \cdot X )^{\Delta}} K_{\Delta - 1}(m \sqrt{X^2})
	\end{equation}
since the light transform only affects the $(-q \cdot X)^{-\Delta}$ piece. Using \eqref{lrs}, we immediately find the light-transformed massive conformal primary wavefunctions
	\begin{equation}
	\widetilde{\Phi}(z, \bz; X) =  \frac{8\pi^2}{m\Gamma(-\Delta) \Gamma(\Delta - 2)}	 (\sqrt{-X^2})^{\Delta-1} \frac{(\d_{z} q \cdot X)^{1-\Delta}}{q \cdot X } K_{\Delta - 1}(m \sqrt{X^2}).
	\end{equation}
The antiholomorphic light transforms can be found using similar methods. 

\subsubsection{Light-ray  OPEs}
\label{Lro}

In this section we show that the light-ray-like exchanges 
\begin{equation}
\label{lrb0}
 \frac{1}{2}\int_{-i\infty}^{i\infty} d\alpha C_{13\alpha}^L C_{24\alpha}^L D^{L,\alpha \alpha} k_{\frac{1}{2} + \alpha}(z) k_{\frac{1}{2} - \alpha}(\bz)
\end{equation}
correspond to light transforms of massive scalars \eqref{light} by matching the OPE coefficients \eqref{OPEs} to OPE coefficients of celestial 3-point functions involving light transforms of massive scalars on the principal series. 
 
 Following \cite{Karateev:2018oml} we write a conformal block associated with an exchange of dimension $\Delta$ and spin $J$ as
 \be 
 k_{\Delta, J}^{\Delta_i}(z, \bz) = \frac{\langle \mathcal{O}_1 \mathcal{O}_3 \mathcal{O}_{\Delta, J}\rangle_0 \langle \mathcal{O}_2 \mathcal{O}_4 \mathcal{O}_{\Delta, J}\rangle_0}{\langle \mathcal{O}_{\Delta, J} \mathcal{O}_{\Delta, J} \rangle_0},
 \ee 
 where $\langle \mathcal{O}_1 \mathcal{O}_2 \mathcal{O}_{\Delta, J}\rangle_0$ and $\langle \mathcal{O}_{\Delta, J} \mathcal{O}_{\Delta, J} \rangle_0$ are stripped conformal three- and two-point structures.
 The conformal block associated with light-ray exchanges then corresponds schematically to the structure 
\begin{equation}
    \frac{\braket{\mathcal{O}_1 L[\mathcal O_{\Delta,J}] \mathcal{O}_3 } \braket{\mathcal{O}_2 L[\mathcal O_{\Delta, J}] \mathcal{O}_4} }{\braket{L[\mathcal O_{\Delta, J}] L[\mathcal O_{\Delta,J}]}}.
\end{equation}
It was shown in \cite{Kravchuk:2018htv} that
\be 
\label{32-pt-lr}
\begin{split}
\langle \mathcal{O}_1 L[\mathcal{O}_{\Delta, 0}] \mathcal{O}_3\rangle &= -2\pi i \frac{1}{{\rm B}\Big(\frac{\Delta + \Delta_{13} }{2}, \frac{\Delta - \Delta_{13}}{2} \Big)} \frac{1}{\Delta - 1} \langle \mathcal{O}_1 \mathcal{O}_3 \mathcal{O}_{1, 1 - \Delta}\rangle_0 \equiv C_{13\alpha}^L, \\
\frac{1}{\langle L[\mathcal{O}_{\Delta, 0}]L[\mathcal{O}_{\Delta, 0}] \rangle} &=-  \frac{\Delta - 1}{2\pi i} \frac{1}{\langle  \mathcal{O}_{1, 1 - \Delta} \mathcal{O}_{1, 1 - \Delta} \rangle}_0 \equiv D^{L, \alpha \alpha}.
\end{split}
\ee 
It then follows that
\begin{equation}
\label{lrb}
\begin{split}
  \frac{\braket{\mathcal{O}_1 L[\mathcal O_{\Delta, 0}]\mathcal{O}_3} \braket{\mathcal{O}_2 L[\mathcal O_{\Delta, 0}] \mathcal{O}_4} }{\braket{L[\mathcal{O}_{\Delta,0}]L[\mathcal{O}_{\Delta,0}]}} = &- \frac{ 2 \pi i}{{\rm B}\Big(\frac{\Delta + \Delta_{13} }{2}, \frac{\Delta - \Delta_{13}}{2} \Big){\rm B}\Big(\frac{\Delta + \Delta_{24} }{2}, \frac{\Delta - \Delta_{24}}{2} \Big) (\Delta - 1)}\\
 &\times k^{\Delta_i}_{1, 1 - \Delta}(z, \bz).
  \end{split}
\end{equation}
A contribution to the massless scalar $4$-point function from exchanges of light-transformed massive scalars with $\Delta = 1 + 2\alpha, \alpha \in i \mathbb{R}$ should then take the form\footnote{The overall factor of $\frac{1}{2}$ also appears in \cite{Karateev:2018oml} in the relation between \eqref{lrb} and the analytic continuation of the conformal block to the Regge regime (cf. equations 5.22, 5.23 therein).}
\be 
\begin{split}
 \frac{1}{2}\int_{-i\infty}^{i\infty} d\alpha D^{\mathcal O \mathcal O} C_{13\mathcal O} C_{24 \mathcal O}& \frac{ -2\pi i}{{\rm B}\Big(\frac{\Delta + \Delta_{13} }{2}, \frac{\Delta - \Delta_{13}}{2} \Big){\rm B}\Big(\frac{\Delta + \Delta_{24} }{2}, \frac{\Delta - \Delta_{24}}{2} \Big) (\Delta - 1)} k^{\Delta_i}_{1, -2\alpha}(z, \bz)\\
&= \frac{1}{2} \int_{-i\infty}^{i\infty} d\alpha C^{L}_{13\alpha} C^L_{24\alpha} D^{L, \alpha\alpha} k_{1,-2\alpha}^{\Delta_i}(z, \bz)
\end{split}
\ee 
and agrees with the block associated with continuous spin in \eqref{res}. In conclusion, the second term in \eqref{res} can be identified with the exchange of light-ray primaries including the expected square of OPE coefficients, up to the  factor of $\csc{\pi\beta \over 2}$.

\section{The contact amplitude}
\label{sec:ca}
Using \eqref{inf-m} and \eqref{res},
we deduce that the 4-point contact amplitude \eqref{cont} has the conformal block representation.
\be 
\label{cbc}
\begin{split}
{\mathcal{A}}_4^c  =  I_{13-24}(z_i, \bz_i) &\frac{2}{\sin \pi \beta/2} \lim_{m \rightarrow \infty}   \left[\frac{1}{2} \int_{-i\infty}^{i\infty} d\alpha c_{13\alpha}^L c_{24\alpha}^L D^{L, \alpha \alpha} 
k_{\frac{1}{2} + \alpha}(z) k_{\frac{1}{2} - \alpha}(\bz)+  \right. \\
&\left. +\sum_{n = 0}^{\infty} c_{13n} c_{24n} D^{n n} \cos\pi\left(h_{13} + \frac{n}{2}\right) \cos\pi\left(h_{24} + \frac{n}{2}\right) k_{\frac{1+n}{2}}(z) k_{\frac{1+n}{2}}(\bz)  \right],
\end{split}
\ee
where 
\be 
\begin{split}
c_{ijn} &=    \sqrt{\lambda} \frac{m^{\Delta_i + \Delta_j - 3}}{2^{\Delta_i +\Delta_j}}  B\left(\frac{1+n}{2} + h_{ij}, \frac{1+n}{2} - h_{ij} \right),\\ c_{ij\alpha}^L &= -\pi i\frac{\sqrt{\lambda}}{m^3}\left({m\over 2}\right)^{2h_i+2h_j}\frac{1}{\alpha},\\
D^{nn} &= m^2 \frac{n}{2 \pi}, \qquad D^{L, \alpha \alpha} = -\frac{ m^2 \alpha^2}{\pi^2 i}.
\end{split}
\ee
These coefficients are consistent with the three-point functions of massless scalars of dimensions $\Delta_1, \Delta_3$ (or $\Delta_2, \Delta_4$) and a massive scalar of mass $m$ and dimension $\Delta = 1 + n$, $n \in \mathbb{Z}_+$.

The decomposition \eqref{cbc} implies that the theory of a massless scalar $\phi$ with a $\phi^4$ interaction alone has no celestial transform: one must  allow for a $\phi^2 \Psi$ coupling to a massive scalar $\Psi$ with mass $m$ which is exchanged  in  conformal blocks. The conformal block decomposition then relates the three-point coupling $g$ to the four-point coupling $\lambda$,
\be 
g = m\sqrt{\lambda/3}.
\ee
Conversely, the tree-level amplitude in $\phi^4$ theory is fully determined by the OPE coefficients of two massless scalars of arbitrary dimensions and a massive scalar of $\Delta = 1 + n,$ with $n$ a positive integer. In appendix \ref{mm} we use the equivalent representation \eqref{res1} of \eqref{res} to compute the conformal block expansion of the tree-level $4$-point scalar amplitude mediated by a scalar of mass $m$ in \eqref{4-sc-sc}. The formulas there at special values of $\beta$ also capture the decompositions of some tree-level and 1-loop 4-point gluon  amplitudes \cite{Gonzalez:2020tpi}.

\section*{Acknowledgements}

We are grateful to Laurent Freidel, Alfredo Guevara, Mina Himwich, Sabrina Pasterski,  Monica Pate, Shu-Heng Shao, Tom Taylor, Pedro Vieira and Xi Yin for useful discussions.  This work was supported in part by DOE grant de-sc/0007870 and the Government of Canada through the Department of Innovation, Science and Industry Canada and by the Province of Ontario through the Ministry of Colleges and Universities. A.A. is supported by NDSEG and Hertz fellowships and A.R. is supported by the Stephen Hawking Postdoctoral Fellowship at Perimeter Institute. 

\appendix

\section{Checking the conformal block decompositions}
\label{cbd}
In this appendix, we perform two nontrivial checks that the conformal block decomposition \eqref{res} is correct.

We first show that for $z, \bar z \ll 1$ the expansion of the four-point contact amplitude for arbitrary external weights correctly reproduces the $z^2 \delta(z-\bar{z})$ behavior of the contact amplitude. The resolution of the 4-point scalar contact celestial amplitude takes the form 
\be
\label{TPCB}
\begin{split}
	&z^2(1-z)^{h_{13}-h_{24}}\delta(z-\bar{z})\\ &=\int_{-i\infty}^{i\infty} \frac{d\alpha}{2\pi i}k_{\frac{1}{2} + \alpha}(z) k_{\frac{1}{2} - \alpha}(\bar{z}) 
	+ \sum_{n = 1}^{\infty} \frac{n}{2\pi^2} c_{13n} c_{24n} \cos\pi \!\left(h_{13} + \frac{n}{2}\right) \cos\pi \!\left(h_{24} + \frac{n}{2}\right)  k_{\frac{1+n}{2}}(z) k_{\frac{1+n}{2}}(\bar{z}) 
\end{split}
\ee
where $c_{13n} c_{24n}$ are the stripped 3-point coefficients  
\be 
\label{opestr}
 \begin{split}
 	c_{13n} &=  B\left(\frac{1}{2}+\frac{n}{2} - h_{13}, \frac{1}{2} +\frac{n}{2}+ h_{13}  \right),\\
 	c_{24n} &=  B\left(\frac{1}{2} + \frac{n}{2} - h_{24}, \frac{1}{2} + \frac{n}{2} + h_{24} \right).
 \end{split}
 \ee
When $z, \bar{z} \ll 1$, the leading term is the integral over continuous $\alpha$, whose contributions have scaling dimension $\Delta = 1$. The remaining blocks have higher scaling dimension and thus contribute higher sub-leading powers in $z$. Using the small-$z$ limit of the conformal blocks, we get
\begin{equation}
\begin{split}
\int_{-i\infty}^{i\infty} \frac{d\alpha}{2\pi i}k_{\frac{1}{2}+\alpha}(z)k_{\frac{1}{2}-\alpha}(\bar{z}) &\to \frac{1}{(2\pi i)}\int_{-i\infty}^{i\infty} d\alpha \sqrt{z\bar{z}}\left(\frac{z}{\bar{z}}\right)^{\alpha}\\
 &= \sqrt{z\bar{z}}\int_{-\infty}^\infty\frac{d\xi}{2\pi}e^{i\xi(\log z - \log \bar{z})} \\
 &= z^2\delta(z-\bar{z}).
\end{split}
\end{equation}
This is precisely the four-point contact amplitude in the small $z$ limit.

We now extend the prior argument to show that the conformal block expansion vanishes when $z \ne \bar{z}$ for generic values of $z, \bar z$ between $0$ and $1$. The continuum term contains an integral over states with $(h, \bar{h}) = (1/2+\alpha, 1/2-\alpha),~ \alpha \in i\rn$.  Because of the growth conditions on these conformal blocks as a function of $\alpha$, for $z = \bar{z}$, the integral cannot be closed and is formally divergent.  For $z > \bar{z}$, the integral can be closed in the left-hand side of the complex plane, while for $z < \bar{z}$ the integral can be closed to the right. Without loss of generality, we focus on the case when $z < \bar{z}$. Then, the continuum term is 
\be
\label{CTSP}
\int_{-i\infty}^{i\infty}\frac{d\alpha}{2\pi i}\sqrt{z\bar{z}}\left(\frac{z}{\bar{z}}\right)^{\alpha}\pFq{2}{1}{\frac12+\alpha-h_{13},\frac12+\alpha+h_{24}}{1+2\alpha}{z}\pFq{2}{1}{\frac12-\alpha-h_{13},\frac12-\alpha+h_{24}}{1-2\alpha}{\bar{z}}.
\ee
For $z < \bar{z}$, we see that integrand is suppressed exponentially at large, positive $\mathrm{Re}(\alpha)$, so the integral can be closed to the right.  The conformal block $k_h(z)$ has a pole when $h$ is a negative integer, so we will find poles at $1-2\alpha = -m, m \in \ints_{\ge 0}$. Using a residue formula for the hypergeometric function 
\begin{equation}
\mathrm{Res}_{c = -m}\left(\pFq{2}{1}{a,b}{c}{z}\right) = \frac{(-1)^mz^{m+1}(a)_{m+1}(b)_{m+1}}{m!(m+1)!}\pFq{2}{1}{1+a+m,1+b+m}{2+m}{z}
\end{equation}
where $(a)_{m+1} = a(a+1)...(a+m)$, \eqref{CTSP} takes the form 
\be
\begin{aligned}
   \frac{1}{(2 \pi i)} &\int_{-i\infty}^{i\infty} d\alpha k_{\frac{1}{2} + \alpha}(z) k_{\frac{1}{2} - \alpha}(\bar{z})  \\
   &= -\sum_{m=0}^\infty k_{1+\frac{m}{2}}(z)\, 
   \bar{z}^{-m/2} \underset{\alpha=\frac{1+m}{2}}{\text{Res}} \left(\pFq{2}{1}{\frac12-\alpha-h_{13},\frac12-\alpha+h_{24}}{1-2\alpha}{\bar{z}}\right) \\
    &= \frac12 \sum_{m=0}^{\infty} k_{1+\frac{m}{2}}(z) k_{1 + \frac{m}{2}} (\bar z)\frac{(-1)^m(-\frac{m}{2}-h_{13})_{m+1}(-\frac{m}{2}+h_{24})_{m+1}}{m!(m+1)!}.  
\end{aligned}
\ee
The factor of $1/2$ arises from
\be 
\lim_{\alpha \rightarrow \frac{1 + m}{2}} \Gamma(1 - 2\alpha) = \frac{(-1)^m}{2 m!(\frac{1+m}{2} - \alpha)}
\ee 
and the sign from the clockwise contour cancels the sign from the residue at the pole in $\alpha$.  We see that the residue of the conformal block $k_{1/2-\alpha}(\bar{z})$ at $\alpha = \frac{1+m}{2}$  reproduces the conformal block $k_{1 + \frac{m}{2}}(z)k_{1 + \frac{m}{2}}(\bz)$. Writing $m + 1 = n$ for $n \in \mathbb{Z}, n \geq 1$ the  coefficient of the conformal block $k_{\frac{1+n}{2}}(z)k_{\frac{1+n}{2}}(\bar{z})$ is
\be
\begin{aligned}
   & - \frac12 (-1)^n n \frac{\Gamma(\frac{1+n}{2}-h_{13}) \Gamma(\frac{1+n}{2}+h_{24})}{\Gamma(\frac{1-n}{2}-h_{13}) \Gamma(\frac{1-n}{2}+h_{24}) (n!)^2 }\\ &=  -  \frac{n}{2 \pi^2} \cos \pi\!\left(\tfrac{n}{2}+h_{13}\right) \cos \pi\!\left(\tfrac{n}{2}+h_{24}\right) \mathrm{B}(\tfrac{1+n}{2}+h_{13}, \tfrac{1+n}{2}-h_{13}) \mathrm{B}(\tfrac{1+n}{2}+h_{24}, \tfrac{1+n}{2}-h_{24}).
\end{aligned}
\ee
This is exactly the negative of the coefficient multiplying the corresponding conformal block in the discrete part of the decomposition. 
Hence, the continuum term generates the $\delta$ function in the four-point contact amplitude, together with a convergent sum which is exactly cancelled by the discrete sum in the conformal block expansion.

\section{Celestial 2-point functions of massive scalars}
\label{m2ptf}

In this appendix we derive the normalization of the celestial two-point function of massive scalars $\chi$ of mass $m$. We start with the momentum space-formula\footnote{We follow the convention of \cite{Lam:2017ofc} where momentum-space amplitudes differ from $S$-matrix elements  by a factor of $i (2\pi)^4.$}
\be 
 (2\pi)^{4}\langle \chi(p_1) \chi(p_2) \rangle =  2p_1^0 (2\pi)^3 \delta^{(3)}(p_1 + p_2),
\ee 
where 
\be 
p_i = m \hat{p}_i, \quad \hat{p}_i^2 = -1.
\ee 
The celestial two-point function is found by integrating against AdS$_3$ bulk-to-boundary propagators \cite{Pasterski:2017kqt},
\be 
\begin{split}
(2\pi)^4 \langle \chi_{\Delta_1}(z_1, \bz_1) \chi_{\Delta_2}(z_2,\bz_2) \rangle &= m^{-2} \int \frac{d^3 \hat{p}_1}{\hat{p}_1^0} \int \frac{d^3 \hat{p}_2}{\hat{p}_2^0} G_{\Delta_1}(\hat{p}_1; z_1, \bz_1) G_{\Delta_2}(\hat{p}_2; z_2, \bz_2) \\
&\times 2\hat{p}_1^0 (2\pi)^3\delta^{(3)}(\hat{p}_1 - \hat{p}_2)\\
&= 2 (2\pi)^3 m^{-2} \int \frac{d^3 \hat{p}_1}{\hat{p}_1^0} G_{\Delta_1}(\hat{p}_1; z_1, \bz_1) G_{\Delta_2}(\hat{p}_1; z_2, \bz_2).
\end{split}
\ee
The orthogonality of bulk-to-boundary propagators 
\be 
\label{ortho}
\begin{split}
&\int \frac{d^3 \hat{p}_1}{\hat{p}_1^0} G_{\Delta_1}(\hat{p}_1; z_1, \bz_1) G_{\Delta_2}(\hat{p}_1; z_2, \bz_2) \\
&= -2\pi^3 \frac{1}{(\Delta_1 - 1)^2}\delta(\Delta_1 + \Delta_2 - 2)\delta^{(2)}(z_1 - z_2) + 2\pi^2\frac{1}{(\Delta_1 - 1)} \delta(\Delta_1 - \Delta_2) \frac{1}{|z_1 - z_2|^{2\Delta_1}}
\end{split}
\ee
then implies that
\be 
\langle \chi_{\Delta_1}(z_1, \bz_1) \chi_{\Delta_2}(z_2,\bz_2) \rangle = \frac{2\pi}{(\Delta_1 - 1) m^2} \delta(\Delta_1 - \Delta_2)\frac{1}{|z_1 - z_2|^{2\Delta_1}} \equiv \delta(\Delta_1 - \Delta_2) \frac{D_{12}}{|z_1 - z_2|^{2\Delta_1}}.
\ee 
We have used the fact that the conformal primary basis for massive particles consists of wavefunctions with $\Delta = 1 + i\lambda$ with $\lambda \geq 0$ \cite{Pasterski:2017kqt} for which the first term on the RHS of \eqref{ortho} vanishes. We can read off
\be 
D_{12} = \frac{2\pi}{(\Delta_1 - 1) m^2} \implies D^{12} = \frac{(\Delta_1 - 1)m^2}{2\pi}
\ee
as promised.

\section{Off-shell exchanges and factorization}
\label{app:t} 

$z$-dependent functions such as \eqref{fz} and \eqref{contact-am} are ubiquitous. Indeed, Poincar\'e symmetry constrains \textit{any} celestial 4-point function to be multiplied by a $\beta$-dependent prefactor. This function arises due to the integral over exchanged off-shell momenta in the bulk Feynman rules. This can be seen from the momentum-space amplitude as follows. Consider the momentum-space 4-point contact amplitude
\begin{equation}
    \mathcal{A}(p_i) = \lambda\delta^{(4)}(p_1 + p_2 + p_3 + p_4).
\end{equation}
We can factor this into a pair of delta functions by integrating over an intermediate off-shell momentum. Due to the kinematic assignment of $p_1$ and $p_2$ ingoing and $p_3$ and $p_4$ outgoing, the intermediate momenta $p$ must have spacelike momentum, so that we integrate over $-p^2 = M^2 \in (-\infty,0]$. 

We now Mellin transform the external legs to arrive at a correlator on the celestial sphere, where we are left with \begin{equation}
\begin{split}
\label{cdec}
  \widetilde{\mathcal{A}} &=  \frac{\lambda}{2} \int_{-\infty}^0 dM^2 M^2 \int_{-i\infty}^{i\infty} d\nu\mu(\nu)\int d^2w \widetilde{A}_3(z_1,z_3,w)\widetilde{A}^*_3(z_2,z_4,w), \\
    \widetilde{A}_3(z_1,z_3,w) &=  \frac{C(\Delta_1,\Delta_3,\Delta)}{|z_{13}|^{\Delta_1+\Delta_3-\Delta}|z_3 - w|^{\Delta +\Delta_3 -\Delta_1}|z_1 - w|^{\Delta_1+\Delta -\Delta_3}}, \\
    C(\Delta_1,\Delta_3,\Delta) &= \frac{M^{\Delta_1+\Delta_3-4}}{2^{\Delta_1+\Delta_3}}B\left(\frac{\Delta_{13}+\Delta}{2},\frac{\Delta_{31}+\Delta}{2}\right), \quad \Delta = 1 + i\nu.
\end{split}
\end{equation}
We can now see how the factor of $\delta(\beta)$ arises. The integrals over $\nu, w$ generate the appropriate $z$ dependence, while the integral over $M^2$, the mass of the exchanged off-shell momentum, becomes 
\begin{equation}
    \int_{-\infty}^{0} dM^2M^{\Delta_1+\Delta_2+\Delta_3+\Delta_4-6} =  2\pi \delta\left(\sum_i h_i-2\right) = 4 \pi \delta(\beta).
\end{equation}
This is precisely the function of $\beta$ present in the 4-point contact amplitude.

We can also show this in the case of a massive scalar exchange. We start with the t-channel scalar amplitude
\begin{equation}
    \mathcal{A}(p_i) = -\frac{g^2}{t - m^2}\delta^{(4)}(p_1 + p_2 + p_3 + p_4).
\end{equation}
As before, this can be written as
\begin{equation}
\begin{split}
    \mathcal{A}(p_i) &= g^2 \int  d^4p \frac{1}{p^2 + m^2}\delta^{(4)}(p_1+p_3-p)\delta^{(4)}(p_2+p_4+p) \\
    &=  g^2 \int d M^2 \int d^4 p \frac{\delta(p^2 + M^2)}{m^2 - M^2} \delta^{(4)}(p_1 + p_3 - p) \delta^{(4)}(p_2 + p_4 + p)\\
    &= \frac{g^2}{2} \int dM^2 M^2 \int\widetilde{d^3\hat{p}}\int \widetilde{d^3\hat{p}'} \frac{1}{m^2 - M^2}\delta^{(4)}(p_1+p_3-p)\delta^{(4)}(p_2+p_4+p') \\
        &\times  \int_{-i\infty}^{i\infty} d\nu\mu(\nu)\int d^2wG_{1+i\nu}(\hat{p};w)G_{1-i\nu}(\hat{p}';w),
\end{split}
\end{equation}
where the integral over $M^2$ runs over the negative real values,
\be 
\widetilde{d^3p} \equiv \frac{d^3 \vec{p}}{p^0}
\ee
and in the last line we used completeness of the bulk-to-boundary propagators,
\be 
\int_{-i\infty}^{i\infty} d\nu\mu(\nu)\int d^2w G_{1+i\nu}(\hat{p};w)G_{1-i\nu}(\hat{p}';w) = \hat{p}^0 \delta^{(3)}(\hat{p} - \hat{p}').
\ee
Just as before, upon taking Mellin transforms with respect to the external lines we find
\be 
\widetilde{A} = \frac{g^2}{2} \int_{-\infty}^0 dM^2 M^2 \frac{1}{m^2 - M^2} \int d\nu \mu(\nu) \int d^2 w \widetilde{A}_3(z_1, z_3, w) \widetilde{A}^*_3(z_2, z_4, w).
\ee
The celestial 3-point functions are given by \eqref{cdec}. We see now that the integral over $M^2$ generates the $1/\sin(\pi \beta/2)$ factor as upon a change of variables and contour integration 
\be 
\int_0^{\infty} dM^2 \frac{M^{\Delta_1 + \Delta_2 + \Delta_3 + \Delta_4 - 6}}{m^2  + M^2}  \propto \frac{\pi m^{\beta - 2}}{\sin (\pi \beta/2)}, \quad \beta = \sum_{i = 1}^4 \Delta_i - 4.
\ee

\section{Amplitudes mediated by massive exchange}

\label{mm}

In this appendix we extend our analysis to 4-point scalar interactions mediated by a massive exchange. For simplicity, we also set $h_{12} = h_{34} = 0$. Taking particles $1, 3$ incoming and $2, 4$ outgoing with dimensions on the principal series, \eqref{4-sc-sc} reduces to 
\be 
\label{4-sc-me}
\widetilde{A}_4^{13 \leftrightarrow 24} = I_{13-24}(z_i, \bar z_i)
\frac{g^2 \pi}{8 m^2} \frac{\left(m/2\right)^{\beta}}{\sin \pi \beta/2}\delta(z - \bz) |z|^{2} \left[ 1 + e^{i \pi\beta/2}|z|^{\beta/2} +  \left|\frac{z}{z - 1} \right|^{\beta/2}\right].
\ee

The first term in \eqref{4-sc-me} is proportional to the 4-point contact amplitude decomposed before so we focus on the remaining two terms. The trick is to replace the prefactor in \eqref{TPCB}  by its conformal block representation \eqref{res} with $h_{13} = h_{24} = 0$ when it multiplies $|z|^{\beta/2}$ and by its equivalent representation 
\be 
\label{res1}
\begin{split}
z^2 \delta(z - \bz) &=  \left[\sum_{k = 0}^{\infty} \frac{2k}{2\pi^2} c_{13(2k)} c_{24(2k)} k_{\frac{1+2k}{2}}\Big(\frac{z}{z - 1}\Big) k_{\frac{1+2k}{2}}\Big(\frac{\bz}{\bz - 1}\Big) \right.\\
&
\left. + \frac{1}{2 \pi i} \int_{-i\infty}^{i\infty} d\alpha k_{\frac{1}{2} + \alpha}\Big(\frac{z}{z - 1}\Big) k_{\frac{1}{2} - \alpha}\Big(\frac{\bz}{\bz - 1}\Big) \right]
\end{split}
\ee
when it multiplies the $\left|\frac{z}{z - 1} \right|^{\beta/2}$ term. \eqref{res1} is obtained by letting $z \rightarrow \dfrac{z}{z - 1}, \bz \rightarrow \dfrac{\bz}{\bz - 1}$ in \eqref{res} and noticing that the LHS is invariant under this change of variables.
$k_h$ are the functions in \eqref{1dcb} with $h_{13} = h_{24} = 0$ and $c_{13(2k)}, c_{24(2k)}$ are \eqref{opestr} evaluated at $h_{13} = h_{24} = 0, n = 2k$. The problem then reduces to finding  the partial wave representations of 
\be 
|z|^{\beta/4} k_{h}(z), \quad \left|\frac{z}{z - 1} \right|^{\beta/4} k_h\left(\frac{z}{z - 1} \right).
\ee
Combining these with their antiholomorphic counterparts and deforming the contour then yields the conformal block decomposition of \eqref{4-sc-me}.
Using the Mellin-Barnes representation of the hypergeometric function \eqref{1dcb}
\be 
k_h(z) = \frac{\Gamma(2h)}{\Gamma(h)^2} \int_{-i\infty}^{i\infty} \frac{ds}{2\pi i} \frac{\Gamma(h + s) \Gamma(h + s) \Gamma(-s)}{\Gamma(2h + s)} z^{h + s} (-1)^s,
\ee
the inner products \cite{Rutter:2019hqc}
\be 
\begin{split}
\langle z^q, \Psi_{\alpha} \rangle &= \frac{\Gamma(q - \frac{1}{2} \pm \alpha)}{\Gamma(q)^2}, \\
\langle \Big(\frac{z}{1 - z} \Big)^q, \Psi_{\alpha} \rangle 
&= \frac{\Gamma(1 - q ) \Gamma(-\frac{1}{2} \pm \alpha + q)}{\Gamma(\frac{1}{2} \pm \alpha)\Gamma(q) }
\end{split}
\ee
and performing the contour integrals, we find
\be 
\begin{split}
\langle z^{\beta/4} k_h(z), \Psi_{\alpha}\rangle &= \frac{\Gamma(h + \beta/4 - \frac{1}{2} \pm \alpha)}{ \Gamma(h + \beta/4)^2} \\
&\times \pFq{4}{3}{h, h, h + \beta/4 - \frac{1}{2} + \alpha, h + \beta/4 - \frac{1}{2} - \alpha}{2h, h + \beta/4, h + \beta/4}{1}.
\end{split}
\ee
Here
\be 
\Gamma(\pm x) \equiv \Gamma(x) \Gamma(-x).
\ee
The lengthier formula for $\langle \left(\frac{z}{1 - z} \right)^{\beta/4} k_{h} \left(\frac{z}{1 - z} \right), \Psi_{\alpha}\rangle$ is given by \eqref{inp2}, \eqref{poles1} below. Plugging these inner products and their antiholomorphic analogs back into \eqref{2Dainv} and deforming the contour in the RH $\alpha,\bar{\alpha}$ planes, we arrive at the conformal block decompositions 
\be 
\begin{split}
|z|^{\beta/2 + 2} \delta(z - \bz)  &=\frac{1}{2 \pi^2}\sum_{n, k, k' = 0}^{\infty} \mathcal{C}_{13(2n)}^k \mathcal{C}_{24(2n)}^{k'}  k_{\frac{2n + 1}{2} + \beta/4 + k}(z) k_{\frac{2n+1}{2} + \beta/4 + k'}(\bz)\\
&+ \frac{1}{2\pi i} \sum_{k, k' = 0}^{\infty} \int_{-i\infty}^{i\infty} d\alpha \mathcal{C}_{13\alpha}^k \mathcal{C}_{24-\alpha}^{k'} k_{\frac{1}{2} + \beta/4 + \alpha + k}(z) k_{\frac{1}{2} + \beta/4 - \alpha + k'}(\bz),
\end{split}
\ee
and 
\be 
\begin{split}
\left|\frac{z}{1 - z}\right|^{\beta/2} |z|^2 & \delta(z - \bz)=
\frac{1}{2 \pi^2}\sum_{n, k, k' = 0}^{\infty} \mathcal{D}_{13(2n)}^k \mathcal{D}_{24(2n)}^{k'}  k_{\frac{2n + 1}{2} + \beta/4 + k}(z) k_{\frac{2n+1}{2}  + \beta/4 + k'}(\bz)\\
&+ \frac{1}{2\pi i} \sum_{k, k' = 0}^{\infty} \int_{-i\infty}^{i\infty} d\alpha \mathcal{D}_{13\alpha}^k \mathcal{D}_{24-\alpha}^{k'} k_{\frac{1}{2} + \alpha  + \beta/4 + k}(z) k_{\frac{1}{2} - \alpha  + \beta/4 + k'}(\bz).
\end{split}
\ee
The OPE coefficients $\mathcal{C}$ and $\mathcal{D}$ are given in \eqref{C} and \eqref{DE}. 
We describe the details on the computation of these two contributions below.

\subsection{$z^{\beta/2}$ term}
\label{zp}

We start off by decomposing
\be 
z^{\beta/4} k_{h}(z)
\ee
in conformal partial waves. We use the Mellin-Barnes representation of the conformal block
\be 
k_h(z) = \underbrace{\frac{\Gamma(2h)}{\Gamma(h)^2}}_{r}\int_{-i\infty}^{i\infty}\frac{ds}{2\pi i} \frac{\Gamma(h + s)^2 \Gamma(-s)}{\Gamma(2h + s)} z^{h + s} (-1)^s
\ee
and the inversion formula of $z^{q}$
\be 
\langle z^q, \Psi_{\alpha} \rangle = \frac{\Gamma(q - \frac{1}{2} \pm \alpha)}{\Gamma(q)^2}.
\ee
Putting everything together, we find
\be 
\begin{split}
\langle z^{\beta/4} k_h(z), \Psi_{\alpha}\rangle &= r \int_{-i\infty}^{i\infty} \frac{ds}{2\pi i} \frac{\Gamma(h + s)^2 \Gamma(-s)}{\Gamma(2h + s)} \frac{\Gamma(h+s +\beta/4 - \frac{1}{2} \pm \alpha)}{\Gamma(h+s + \beta/4)^2} (-1)^s.
\end{split}
\ee
We now close the contour in the RHP to find
\be 
\begin{split}
\langle z^{\beta/4} k_h(z), \Psi_{\alpha}\rangle &= \frac{\Gamma(h + \beta/4 - \frac{1}{2} \pm \alpha)}{ \Gamma(h + \beta/4)^2} \\
&\times \pFq{4}{3}{h, h, h + \beta/4 - \frac{1}{2} + \alpha, h + \beta/4 - \frac{1}{2} - \alpha}{2h, h + \beta/4, h + \beta/4}{1}.
\end{split}
\ee

Notice that the only poles in $\alpha$ come from $\Gamma(h + \frac{\beta}{4} - \frac{1}{2} \pm \alpha)$. Moreover \eqref{res} is symmetric in $z \leftrightarrow \bz$ so the alpha space transform will be symmetric under $\alpha \leftrightarrow \bar{\alpha}.$ We find
\be 
z^{\beta/2 + 2} \delta(z - \bz) = \int_{-i\infty}^{i\infty} \frac{d\alpha d \bar{\alpha}}{Q(-\alpha) Q(-\bar{\alpha})} f(\alpha, \bar{\alpha}) k_{\frac{1}{2} + \alpha}(z) k_{\frac{1}{2} + \bar{\alpha}}(\bz),
\ee
with 
\be 
\begin{split}
f(\alpha, \bar{\alpha}) &= \frac{1}{2\pi^2}\sum_{n = 0}^{\infty} 2n c_{13(2n)} c_{24(2n)} \langle z^{\beta/4} k_{\frac{2n + 1}{2}}(z), \Psi_{\frac{1}{2} + \alpha}\rangle \langle z^{\beta/4} k_{\frac{2n + 1}{2}}(\bz), \Psi_{\frac{1}{2} + \bar{\alpha}}\rangle \\
&+ \frac{1}{(2\pi i)} \int_{-i\infty}^{i\infty} d\alpha' \langle z^{\beta/4} k_{\frac{1}{2} + \alpha'}(z), \Psi_{\frac{1}{2} + \alpha}\rangle \langle z^{\beta/4} k_{\frac{ 1}{2} - \alpha'}(\bz), \Psi_{\frac{1}{2} + \bar{\alpha}}\rangle.
\end{split}
\ee

In both terms we can deform the contour in the RH $\alpha, \bar{\alpha}$ planes to only pick up poles at $\alpha = h + \frac{\beta}{4} - \frac{1}{2} + k$ with $h = \frac{1+ 2n}{2}, ~\frac{1}{2} + \alpha'$ respectively (and similar equations for $\bar{\alpha}$).  We conclude that 
\be 
\begin{split}
z^{\beta/2 + 2} \delta(z - \bz)  &=\frac{1}{2\pi^2}\sum_{n, k, k' = 0}^{\infty}2n \mathcal{C}_{13(2n)}^k \mathcal{C}_{24(2n)}^{k'} k_{\frac{2n + 1}{2} + \frac{\beta}{4} + k}(z) k_{\frac{2n+1}{2} + \frac{\beta}{4} + k'}(\bz)\\
&+ \frac{1}{2\pi i} \sum_{k, k' = 0}^{\infty} \int_{-i\infty}^{i\infty} d\alpha \mathcal{C}_{13\alpha}^k \mathcal{C}_{24-\alpha}^{k'} k_{\frac{1}{2} + \alpha + \frac{\beta}{4} + k}(z) k_{\frac{1}{2} - \alpha + \frac{\beta}{4} + k'}(\bz),
\end{split}
\ee
where 
\be 
\label{C}
\begin{split}
\mathcal{C}_{13n}^k &= c_{13n} \frac{(-1)^k}{k!} \frac{\Gamma(n + \beta/2 + k)}{ \Gamma(\frac{n + 1}{2} + \frac{\beta}{4})^2} \pFq{4}{3}{\frac{n + 1}{2}, \frac{n + 1}{2}, n + \beta/2 + k, -k}{n+1, \frac{n+1}{2} + \frac{\beta}{4}, \frac{n+1}{2} + \frac{\beta}{4} }{1} \frac{1}{Q(-\frac{n}{2} - \frac{\beta}{4}  - k)},\\
\mathcal{C}_{13\alpha}^k &= \frac{(-1)^k}{k!}\frac{\Gamma( 2\alpha  + \beta/2 +k)}{\Gamma(\frac{1}{2} + \alpha + \frac{\beta}{4})\Gamma(\frac{1}{2} + \alpha + \frac{\beta}{4})}\\ &\times \pFq{4}{3}{\frac{1}{2} + \alpha, \frac{1}{2} + \alpha,  2 \alpha + \beta/2  + k, -k }{2\alpha + 1, \frac{1}{2} + \alpha + \frac{\beta}{4}, \frac{1}{2} + \alpha + \frac{\beta}{4}}{1} \frac{1}{Q( - \alpha - \frac{\beta}{4}  - k)}.
\end{split}
\ee

It is not hard to see that for $\beta = 0,$ the generalized hypergeometric function collapses to a ratio of gamma functions which vanishes for $k> 0$ as expected. Note that the discrete tower of exchanges now appear at positive integer dimensions shifted by $\beta/2$, consistent with the behavior of the celestial 4-point amplitude in the small $z$ limit.

\subsection{$\left|\dfrac{z}{z - 1}\right|^{\beta/2}$ term}

We begin by computing
\be 
\begin{split}
\langle \Big(\frac{z}{1 - z} \Big)^{q}, \Psi_{\alpha} \rangle &= \int_0^1 \frac{dz}{z^2} \Big(\frac{z}{1-z} \Big)^{q} \frac{1}{\Gamma(\frac{1}{2} \pm \alpha)} \int \frac{ds}{2\pi i} \frac{\Gamma(-s)\Gamma(\frac{1}{2} + s \pm \alpha)}{\Gamma(1 + s)} \left(\frac{1 - z}{z} \right)^s\\
&= \int \frac{ds}{2\pi i}  \frac{\Gamma(-s)\Gamma(\frac{1}{2} + s \pm \alpha)\Gamma(q - s  - 1)\Gamma(- q + s + 1)}{\Gamma(1 + s) \Gamma(\epsilon)} \frac{1}{\Gamma(\frac{1}{2} \pm \alpha)}\\
&= \frac{\Gamma(\frac{1}{2} \pm \alpha)\Gamma(- q + 1) \Gamma(-\frac{1}{2} \pm \alpha + q)}{\Gamma(\frac{1}{2} \pm \alpha)\Gamma(q)}\frac{1}{\Gamma(\frac{1}{2}\pm \alpha)}\\
&= \frac{\Gamma(- q + 1) \Gamma(-\frac{1}{2} \pm \alpha + q)}{\Gamma(\frac{1}{2} \pm \alpha)\Gamma(q) }.
\end{split}
\ee
We can now repeat the analysis of section \ref{zp} using the alternate form of the delta function resolution \eqref{res1}. We first notice that
\be 
\label{inp2}
\begin{split}
&\langle \left|\frac{z}{z - 1} \right|^{\beta/4} k_{h}\left(\frac{z}{z - 1} \right), \Psi_{\alpha}\rangle = (-1)^{h} r \int_{-i\infty}^{i\infty} \frac{ds}{2\pi i}  \frac{\Gamma(h + s)^2 \Gamma(-s)}{\Gamma(2h + s)} \langle\left( \frac{z}{1 - z} \right)^{\beta/4 + h + s}, \Psi_{\alpha} \rangle \\
&= r  (-1)^h\int_{-i\infty}^{i\infty} \frac{ds}{2\pi i}  \frac{\Gamma(- (\beta/4 + h + s) + 1) \Gamma(-\frac{1}{2} \pm \alpha + \beta/4 + h + s)}{\Gamma(\frac{1}{2} \pm \alpha)\Gamma(\beta/4 + h + s) }\frac{\Gamma(h + s)^2 \Gamma(-s)}{\Gamma(2h + s)}.\\
\end{split}
\ee

This formula has two pole sequences at
\be 
s = k, \qquad s = -h - \beta/4 + 1 + k.
\ee
The first gives
\be 
\label{poles1}
\begin{split}
&(-1)^h\frac{ \Gamma(-\frac{1}{2} \pm \alpha + \frac{\beta}{4} + h) \Gamma(1 - h - \beta/4)}{\Gamma(\frac{\beta}{4} + h) \Gamma
(\frac{1}{2} \pm \alpha)} \pFq{4}{3}{h, h , -\frac{1}{2} \pm \alpha + \frac{\beta}{4} + h}{\frac{\beta}{4} + h, 2h, h + \frac{\beta}{4}}{1},
\end{split}
\ee
while the second yields a term with no poles in $\alpha,$ hence doesn't contribute to the conformal block decomposition. 
As before, we can close the contour in the RHS $\alpha, \bar{\alpha}$ planes picking up poles at
\be 
\alpha = \frac{\beta - 2}{4} + h + k,
\ee
with $h = \frac{1}{2} + n, \frac{1}{2} + \alpha'$ respectively and similarly for $\bar{\alpha}$. The conformal block decomposition looks similar to that in the previous section, namely

\be 
\begin{split}
\left|\frac{z}{1 - z}\right|^{\beta/2} z^2 \delta(z - \bz) &=
\frac{1}{2\pi^2}\sum_{n, k, k' = 0}^{\infty}  2n\mathcal{D}_{13(2n)}^k \mathcal{D}_{24(2n)}^{k'} k_{\frac{2n + 1}{2} + \frac{\beta}{4} + k}(z) k_{\frac{2n+1}{2} + \frac{\beta}{4} + k'}(\bz)\\
&+ \frac{1}{2\pi i} \sum_{k, k' = 0}^{\infty} \int_{-i\infty}^{i\infty} d\alpha \mathcal{D}_{13\alpha}^k \mathcal{D}_{24-\alpha}^{k'} k_{\frac{1}{2} + \alpha + \frac{\beta}{4} + k}(z) k_{\frac{1}{2} - \alpha + \frac{\beta}{4} + k'}(\bz),
\end{split}
\ee
with
\be 
\label{DE}
\begin{split}
\mathcal{D}_{13n}^k &= c_{13n} \frac{(-1)^{\frac{n + 1}{2} + k}}{k!} \frac{\Gamma(n + \beta/2 + k) \Gamma(-\beta/4 - \frac{n-1}{2})}{\Gamma(\frac{\beta}{4} + \frac{n + 1}{2}) \Gamma
(\frac{1}{2} \pm (\frac{2n + \beta}{4} +k)) }\\
&\times \pFq{4}{3}{\frac{n + 1}{2}, \frac{n+1}{2}, k + \beta/2 + n, -k}{\frac{\beta}{4} + \frac{n + 1}{2}, n+1, \frac{n + 1}{2} + \frac{\beta}{4}}{1} \frac{1}{Q(-\frac{\beta}{4} - \frac{n}{2} - k)}
\end{split}
\ee
and an analogous expression for $\mathcal{D}_{13\alpha}^k$.
The exchanges are again at $\Delta \sim  \beta/2 + k + k', J \sim k - k'$ (which is what we'd get if the amplitude wasn't multiplied by $\delta(z - \bz)$) together with a tower of exchanges with $\Delta$ shifted by positive even integers as a consequence of translation invariance.

\end{document}